\documentclass[aps,prl,showpacs,twocolumn,preprintnumbers,superscriptaddress,nofootinbib,floatfix,10pt]{revtex4-1}
\usepackage{amsfonts,amssymb,stmaryrd,latexsym,amsmath,braket}
\usepackage{graphicx,subfigure}
\usepackage{comment}
\usepackage{times}
\usepackage{slashed}
\usepackage{bm}
\usepackage{braket}
\usepackage{chapterbib}
\usepackage{color}

\newcommand{\beginsupplement}{%
        \setcounter{table}{0}
        \renewcommand{\thetable}{S\arabic{table}}%
        \setcounter{figure}{0}
        \renewcommand{\thefigure}{S\arabic{figure}}%
        \setcounter{equation}{0}
        \renewcommand{\theequation}{S\arabic{equation}}%
     }

\makeatletter
\let\saved@includegraphics\includegraphics
\AtBeginDocument{\let\includegraphics\saved@includegraphics}
\makeatother

\begin{document}

\title{{\em Ab initio} study of the radii of oxygen isotopes}


\author{Zhengxue Ren}
\email{zxren@nankai.edu.cn}
\affiliation{Institute for Advanced Simulation (IAS-4), Forschungszentrum J\"{u}lich, D-52425 J\"{u}lich, Germany}
\affiliation{Helmholtz-Institut f\"{u}r Strahlen- und Kernphysik and Bethe Center for Theoretical Physics, Universit\"{a}t Bonn, D-53115 Bonn, Germany}
\affiliation{School of Physics, Nankai University, Tianjin 300071, China}

\author{Serdar Elhatisari}
\email{selhatisari@gmail.com}
\affiliation{King Fahd University of Petroleum and Minerals (KFUPM), 31261 Dhahran, Saudi Arabia}
\affiliation{Faculty of Natural Sciences and Engineering, Gaziantep Islam Science and Technology University, Gaziantep 27010, Turkey}
%\affiliation{Helmholtz-Institut f\"{u}r Strahlen- und Kernphysik and Bethe Center for Theoretical Physics, Universit\"{a}t Bonn, D-53115 Bonn, Germany}

\author{Ulf-G.~Mei{\ss}ner}
\email{meissner@hiskp.uni-bonn.de}
\affiliation{Helmholtz-Institut f\"{u}r Strahlen- und Kernphysik and Bethe Center for Theoretical Physics, Universit\"{a}t Bonn, D-53115 Bonn, Germany}
\affiliation{Institute for Advanced Simulation (IAS-4), Forschungszentrum J\"{u}lich, D-52425 J\"{u}lich, Germany}
\affiliation{Peng Huanwu Collaborative Center for Research and Education, International Institute for Interdisciplinary and Frontiers, Beihang University, Beijing 100191, China}

\begin{abstract}
We present an {\em ab initio} study of the charge and matter radii of oxygen isotopes from $^{16}$O to $^{20}$O using nuclear lattice effective field theory (NLEFT) with high-fidelity N$^3$LO chiral interactions. To efficiently address the Monte Carlo sign problem encountered in nuclear radius calculations, we introduce the {\em partial pinhole algorithm}, significantly reducing statistical uncertainties and extending the reach to more neutron-rich and proton-rich isotopes. Our computed charge radii for $^{16}$O, $^{17}$O, and $^{18}$O closely match experimental data, and we predict a charge radius of $2.810(32)$~fm for $^{20}$O. The calculated matter radii show excellent agreement with values extracted from low-energy proton and electron elastic scattering data, but are inconsistent with those derived from interaction cross sections and charge-changing cross section measurements. These discrepancies highlight model-dependent ambiguities in the experimental extraction methods of matter radii and underscore the value of precise theoretical benchmarks from NLEFT calculations.
\end{abstract}
\maketitle
\date{today}

%%%%%%%%%%%%%%%%%%%%%%%%%%%%%%%%%%%%%%%%%%%%%%%%%%%%%%%%%%%%%%%%%%%%%
% introduction
\section{Introduction}
The nuclear radius, which characterizes the size of the nucleus, is one of the most fundamental properties of atomic nuclei.
With the advancement of experimental techniques, there has been a rapid increase in available data on nuclear charge and matter radii for both stable and unstable isotopes~\cite{Ozawa:2000gx, Angeli:2013epw, Campbell:2016yxe, Yang:2022wbl}.
On the one hand, these data have significantly enhanced our understanding of nuclear structure, revealing novel phenomena such as nuclear halos~\cite{Tanihata:1985psr, Kanungo:2016tmz}, neutron skins~\cite{PREX:2021umo, CREX:2022kgg}, and nuclear shape phase transitions~\cite{Marsh:2018wxs, Cubiss:2023cyb}.
On the other hand, they pose considerable challenges to nuclear theory, particularly to {\em ab initio} approaches.

Despite the remarkable progress in nuclear theory, achieving accurate descriptions of nuclear charge radii remains a significant challenge~\cite{Cipollone:2014hfa, GarciaRuiz:2016ohj, Lapoux:2016exf, Miller:2019proton, Koszorus:2020mgn, Kaur:2022yoh, Konig:2023bzp}.
A notable example is the oxygen isotope chain.
Most {\em ab initio} calculations underestimate the charge radius of $^{16}$O by roughly 7--15\%~\cite{Cipollone:2014hfa, Ekstrom:2015rta}.
Furthermore, while $^{16}$O and $^{17}$O exhibit similar charge radii, a pronounced increase in the charge radius is observed experimentally for $^{18}$O in comparison to $^{16}$O and $^{17}$O~\cite{Angeli:2013epw}.
To our knowledge, this intriguing feature is not captured by any existing theoretical approach.
Although the chiral interaction ${\rm NNLO}_{\rm sat}$ successfully reproduces the charge radius of $^{16}$O by including it among the fitted observables~\cite{Ekstrom:2015rta},
it still fails to give the substantial increase seen in $^{18}$O~\cite{Lapoux:2016exf, Kaur:2022yoh}.

Nuclear lattice effective field theory (NLEFT)~\cite{Lee:2008fa, Lahde2019book} offers a different perspective on this problem.
Recently, the wave function matching method~\cite{Elhatisari:2022zrb} was proposed to mitigate the notorious Monte Carlo sign problem in the imaginary-time evolution of quantum many-body simulations,
which introduces strong cancellations between positive and negative amplitudes.
In combination with high-fidelity N$^3$LO chiral interactions, the binding energies of both light and medium-mass nuclei up to $A=58$ are well reproduced.
While charge radii are generally well described, including the pronounced increase in $^{18}$O compared to $^{16,17}$O, they are often accompanied by significantly larger statistical uncertainties.
This is largely due to the commonly used pinhole algorithm, which inserts an $A$-body density operator during the imaginary-time evolution~\cite{Elhatisari:2017eno}.
This insertion leads to unpaired nucleons which results in additional strong sign-oscillations, particularly for nuclei with larger mass numbers and/or pronounced proton–neutron asymmetry, thereby complicating the investigation of subtle isotopic trends in nuclear radii.

In this work, we propose an extension of the pinhole algorithm, referred to as the {\em partial pinhole algorithm}, that reduces the rank of the many-body density operator in order to mitigate the sign problem in nuclear radius calculations within the framework of NLEFT.
This reduced operator is implemented via the recently developed rank-operator method~\cite{Ma:2023ahg}.
We combine the  partial pinhole algorithm with the wave function matching method using high-fidelity N$^3$LO chiral interactions.
As a first application, we investigate both charge and matter radii of a number of oxygen isotopes.

\section{Formalism}
The full details of our lattice calculations are shown in the Supplemental Material~\cite{SM}.
To illustrate the idea of the partial pinhole algorithm and how it is combined with NLEFT in calculating nuclear radii,
we adopt a simplified Hamiltonian with an attractive two-body contact short range interaction ($C<0$),
\begin{equation}
  H=K+\frac{1}{2}C\,\sum_{\bm{n}}:\rho^2(\bm{n}):,
\end{equation}
where $K$ is the kinetic energy term with nucleon mass $m = 938.9$~MeV, the colons indicate normal ordering, and $\rho(\bm{n})=\sum_{\sigma\tau}\rho_{\sigma\tau}(\bm{n})=\sum_{\sigma\tau}a^\dagger_{\sigma\tau}(\bm{n})a_{\sigma\tau}(\bm{n})$.
Here, $\rho_{\sigma\tau}(\bm{n})$ is the one-body density operator at lattice site $\bm{n}$ for spin $\sigma$ and isospin $\tau$.

The ground state wave functions of $H$ can be obtained by applying imaginary time projectors to a trial wave function $\Psi_0$, $|\Psi\rangle=\lim_{\tau\to\infty}e^{-H\tau/2}|\Psi_0\rangle$.
In Monte Carlo simulations, $\tau$ is divided into $L_t$ slices with temporal spacing $a_t$ such that $\tau= L_ta_t$.
For each time slice, the two-body interaction is defined as nucleons propagated in a fluctuating background auxiliary field using a Hubbard-Stratonovich transformation,
\begin{equation}
  \exp\left(-\frac{a_tC}{2}\rho^2\right) = \sqrt{\frac{1}{2\pi}} \int ds~\exp\left(-\frac{s^2}{2}+\sqrt{-a_tC}s\rho\right),
\end{equation}
where $s$ is the auxiliary field at a lattice site.
Therefore, the wave function $|\Psi\rangle$ can be written as an auxiliary field path-integral expression,
\begin{equation}\label{Eq:wfs}
   |\Psi\rangle = \int \mathcal{D}{s_1}\cdots \mathcal{D}{s_{L_t/2}} M(s_{Lt/2})\cdots M(s_1) |\Psi_0\rangle~,
\end{equation}
where the initial wave function $|\Psi_0\rangle$ is a $A$-nucleon Slater determinant, such as alpha clusters states or shell-model wave functions, and the normal-ordered transfer matrix $M$ is defined as,
\begin{equation}
    M(s_{n_t}) = :\exp\left[-a_tK+\sqrt{-a_tC}\sum_{\bm{n}}s_{n_t}(\bm{n})\rho(\bm{n})\right]:.
\end{equation}
Therefore, $|\Psi\rangle$ is expressed as a linear combination of $A$-nucleon Slater determinants, each of which is associated with a specific configuration of auxiliary fields $\vec{s}$.

With the wave function $|\Psi\rangle$ in Eq.~\eqref{Eq:wfs}, the root-mean-square (RMS) point-proton radius of a nucleus with proton number $Z$, neutron number $N$, and mass number $A$ can be evaluated by
\begin{equation}
  \langle r_p^2\rangle = \frac{1}{Z} \sum_{\bm{n},\sigma,\tau=p} \langle \rho_{\sigma\tau}(\bm{n}) (\bm{n} - \hat{\bm{r}}_{\rm c.m.})^2 \rangle,
\end{equation}
where $\hat{\bm{r}}_{\rm c.m.} = \sum_{\bm{n},\sigma,\tau} \bm{n} \rho_{\sigma\tau}(\bm{n}) / A$ is the center-of-mass operator.
The expression can be rewritten in terms of a set of two-body correlation functions $G_{\alpha\beta}$ as
\begin{equation}
  \langle r_p^2\rangle = \frac{N}{ZA^2} G_{pp} + \frac{N-Z}{ZA^2} G_{pn} - \frac{1}{A^2} G_{nn}~,
\end{equation}
with
\begin{equation}\label{eq:Gab}
  G_{\alpha\beta} = \sum_{\substack{\bm{n}_1,\sigma_1,\tau_1 = \alpha \\ \bm{n}_2,\sigma_2,\tau_2 = \beta}} \langle :\rho_{\sigma_1\tau_1}(\bm{n}_1) \rho_{\sigma_2\tau_2}(\bm{n}_2): \rangle (\bm{n}_1 - \bm{n}_2)^2.
\end{equation}
Similarly, the matter radius is given by
\begin{align}
  \langle r_m^2\rangle &= \frac{1}{A} \sum_{\bm{n},\sigma,\tau} \langle \rho_{\sigma\tau}(\bm{n}) (\bm{n} - \hat{\bm{r}}_{\rm c.m.})^2 \rangle \nonumber \\
  &= \frac{1}{2A^2} G_{pp} + \frac{1}{A^2} G_{pn} + \frac{1}{2A^2} G_{nn}.
\end{align}
Thus, nuclear radii can be expressed in terms of just three two-body correlation functions. It is important to note that since the matter radius corresponds to the second radial moment of the intrinsic nuclear density distribution (or equivalently, the squared nuclear wave function), its numerical value is sensitive to both the theoretical framework used and the specific methods adopted when extracting it from experimental observables.

Similar to the previous studies on computing the RMS charge radii~\cite{Ekstrom:2015rta,Elhatisari:2022zrb, Sun:2025yfo}, we use the standard relation~\cite{Friar:1975Advances}
\begin{equation}
  r_{\rm ch}^2 = \langle r_p^2 \rangle + R_p^2 + \frac{N}{Z} R_n^2 + \frac{3}{4m_p^2},
\end{equation}
where $R_p^2 = 0.7056~{\rm fm}^2$~\cite{Pohl:2010zza, Lin:2021xrc}, $R_n^2 = -0.105~{\rm fm}^2$~\cite{Filin:2020tcs}, and $m_p = 938.27$~MeV.
Relativistic spin-orbit corrections and two-nucleon current contributions are not included in this work.

A direct calculation of the correlation functions $G_{\alpha\beta}$ is computationally expensive,
especially when the corrections from high-fidelity chiral interactions are included perturbatively.
To evaluate them more efficiently, we introduce the \emph{partial pinhole algorithm}.
We define a normal-ordered $M$-body density operator,
\begin{equation}\label{eq:rhoM}
  \rho_M(\vec{c}\,) = :\rho_{\sigma_1\tau_1}(\bm{n}_1) \cdots \rho_{\sigma_M\tau_M}(\bm{n}_M):,
\end{equation}
using the notation $\vec{c} = (c_1, \cdots, c_M)$.
Here, $c_i = (\bm{n}_i, \sigma_i, \tau_i)$ specifies the quantum numbers of the $i$-th density operator, with $\bm{n}_i$ denoting the lattice coordinate, and $\sigma_i$ and $\tau_i$ representing spin and isospin.
The summation over $\vec{c}$ satisfies
\begin{align}
  \sum_{\vec{c}} \rho_M(\vec{c}\,) &= \hat{N}(\hat{N} - 1) \cdots (\hat{N} - M + 1) \nonumber \\
  &= A(A - 1) \cdots (A - M + 1),
\end{align}
where $\hat{N}$ is the particle-number operator.
The equality of the second line stems from the fact that we are working in $A$-nucleon subspace.
Therefore, this summation is proportional to the identity operator.

The correlation functions $G_{\alpha\beta}$ can then be evaluated as
\begin{align}\label{Eq:corr_G}
  G_{\alpha\beta} &= \frac{A(A - 1)}{M(M - 1)} \nonumber \\
  &\quad \times \frac{\displaystyle\sum_{\vec{c}} \left[ \langle \Psi | \rho_M(\vec{c}\,) | \Psi \rangle \sum_{i<j} \delta_{\tau_i\alpha} \delta_{\tau_j\beta} (\bm{n}_i - \bm{n}_j)^2 \right]}{ \displaystyle\sum_{\vec{c}} \langle \Psi | \rho_M(\vec{c}\,) | \Psi \rangle }.
\end{align}
The details about computing $(\bm{n}_i - \bm{n}_j)^2$ are presented in the Supplemental Material~\cite{SM}.
Unlike the standard pinhole algorithm~\cite{Elhatisari:2017eno}, the value of $M$ is much smaller than the total number of nucleons $A$.
As a result, $\rho_M$ is no longer a projection operator, and acting with $\rho_M$ on a single $A$-nucleon Slater determinant yields a linear combination of multiple $A$-nucleon Slater determinants.

Since $\rho_{\sigma\tau}(\bm{n})$ is a rank-one operator, that is, $:\rho_{\sigma\tau}^2(\bm{n}): = 0$, the rank-one operator method~\cite{Ma:2023ahg} allows us to express $\rho_M(\vec{c}\,)$ as
\begin{equation}
  \rho_M(\vec{c}\,) = \frac{1}{\varepsilon^M} :\exp[ \varepsilon \rho_{\sigma_1\tau_1}(\bm{n}_1) + \cdots + \varepsilon \rho_{\sigma_M\tau_M}(\bm{n}_M) ]:
\end{equation}
with $\varepsilon \to \infty$.
It consists of a string of one-body operators which can be applied directly to each single-particle wave function in the Slater determinant.

The summations over $\vec{c}$ in Eq.\eqref{Eq:corr_G} and the path integral over $\vec{s}$ in Eq.\eqref{Eq:wfs} are evaluated via Monte Carlo importance sampling. We generate an ensemble of ${\vec{c},\vec{s}}$ configurations according to the relative probability weight,
\begin{align}
  P(\vec{c},\vec{s}\,) = |\langle \Psi_0 |&M(s_{L_t})\cdots M(s_{L_t/2+1})\rho_M(\vec{c}\,)\nonumber\\
   &\times M(s_{L_t/2})\cdots M(s_1)| \Psi_0 \rangle|.
\end{align}
The strings of transfer matrices are generated by updating the auxiliary fields $\vec{s}$ using the shuttle algorithm~\cite{Lu:2018bat}. At the middle time slice, the auxiliary field updates are paused, and the pinhole configuration $\vec{c}$ is updated by performing standard Metropolis accept/reject steps. Specifically, to update $\vec{c}$, we randomly select an index $i$ in $\rho_M$ and either move it to a neighboring site,
\begin{equation}
  c_i = \{\bm{n}_i, \sigma_i, \tau_i\} \rightarrow c_i' = \{\bm{n}_i', \sigma_i, \tau_i\},
\end{equation}
or reassign its spin and isospin,
\begin{equation}
  c_i = \{\bm{n}_i, \sigma_i, \tau_i\} \rightarrow c_i' = \{\bm{n}_i, \sigma_i', \tau_i'\}.
\end{equation}
The new configuration $\vec{c}\,'$ is accepted if
\begin{equation}
  { P(\vec{c}\,',\vec{s}\,)}/{ P(\vec{c},\vec{s}\,)} > r,
\end{equation}
where the random number $r$ is uniformly distributed between 0 and 1.

In this work, we combine the partial pinhole algorithm with the wave function matching method.
The original chiral Hamiltonian is unitarily transformed into a new high-fidelity Hamiltonian $H$, whose wave function matches that of a computationally simple Hamiltonian, $H_S$,  at a given radius.
This ensures rapid convergence of the perturbative expansion in $H - H_S$.
The details can be found in Ref.~\cite{Elhatisari:2022zrb}.
The formalism for applying the partial pinhole algorithm in perturbative calculations is provided in~\cite{SM}.

\section{Results and discussion}

In the following calculations, we adopt the N$^3$LO chiral interactions from Ref.~\cite{Elhatisari:2022zrb}.
We also use a  minimal sets of three-body force terms, where, in addition to nuclear binding energies, the charge radius of $^4$He is used to determine the 
low-energy constants (LECs)~\cite{Elhatisari:pre}.
Although the number of involved free LECs is reduced from eight to six, the overall description of nuclear binding energies is similar to Ref.~\cite{Elhatisari:2022zrb}, as detailed in~\cite{SM}.
Our simulations are performed with a spatial lattice spacing of $a = 1.32~{\rm fm}$, corresponding to a momentum cutoff $\Lambda = \pi/a \approx 471~{\rm MeV}$. Additionally, we use a temporal lattice spacing of $a_t = 0.001~{\rm MeV}^{-1}$ and a cubic box of length $L = 13.2~{\rm fm}$. Since the radii of the oxygen isotopes under investigation are all below 3~fm, this box size is clearly sufficient to suppress finite-volume effects.

In the partial pinhole algorithm, although one could in principle take $M=2$ for $\rho_M$ in Eq.~\eqref{eq:rhoM} as the correlation functions $G_{\alpha\beta}$ in Eq.~\eqref{eq:Gab} are two-body observables, such a choice suffers from low computational efficiency.
To improve sampling efficiency, we aim for each configuration $\vec{c}$ to contribute simultaneously to all three types of correlations: $G_{pp}$, $G_{pn}$, and $G_{nn}$.
This requirement implies that the isospin indices in $\rho_M$ must include at least two protons and two neutrons.
We therefore adopt $M=4$ and fix half of the indices to be protons and the other half to be neutrons in our calculations.
We have further verified that the final results are largely insensitive to the specific choice of $M$, having tested $M=4$, 6, 8, and 10~\cite{SM}.

In~\cite{SM}, we benchmark the charge radii of $^{16}$O and $^{17}$O using both the standard pinhole and the partial pinhole algorithms with the simple Hamiltonian $H_S$.
Although we find that both methods yield consistent charge radii, the partial pinhole algorithm maintains a significantly larger average phase factor and substantially reduces statistical uncertainties, particularly in the case of $^{17}$O.
This improvement becomes even more substantial for heavier nuclei such as $^{40}$Ca,
and is further amplified when applying the full N$^3$LO chiral interactions.

\begin{figure}[!htbp]
  \centering
  \includegraphics[width=0.45\textwidth]{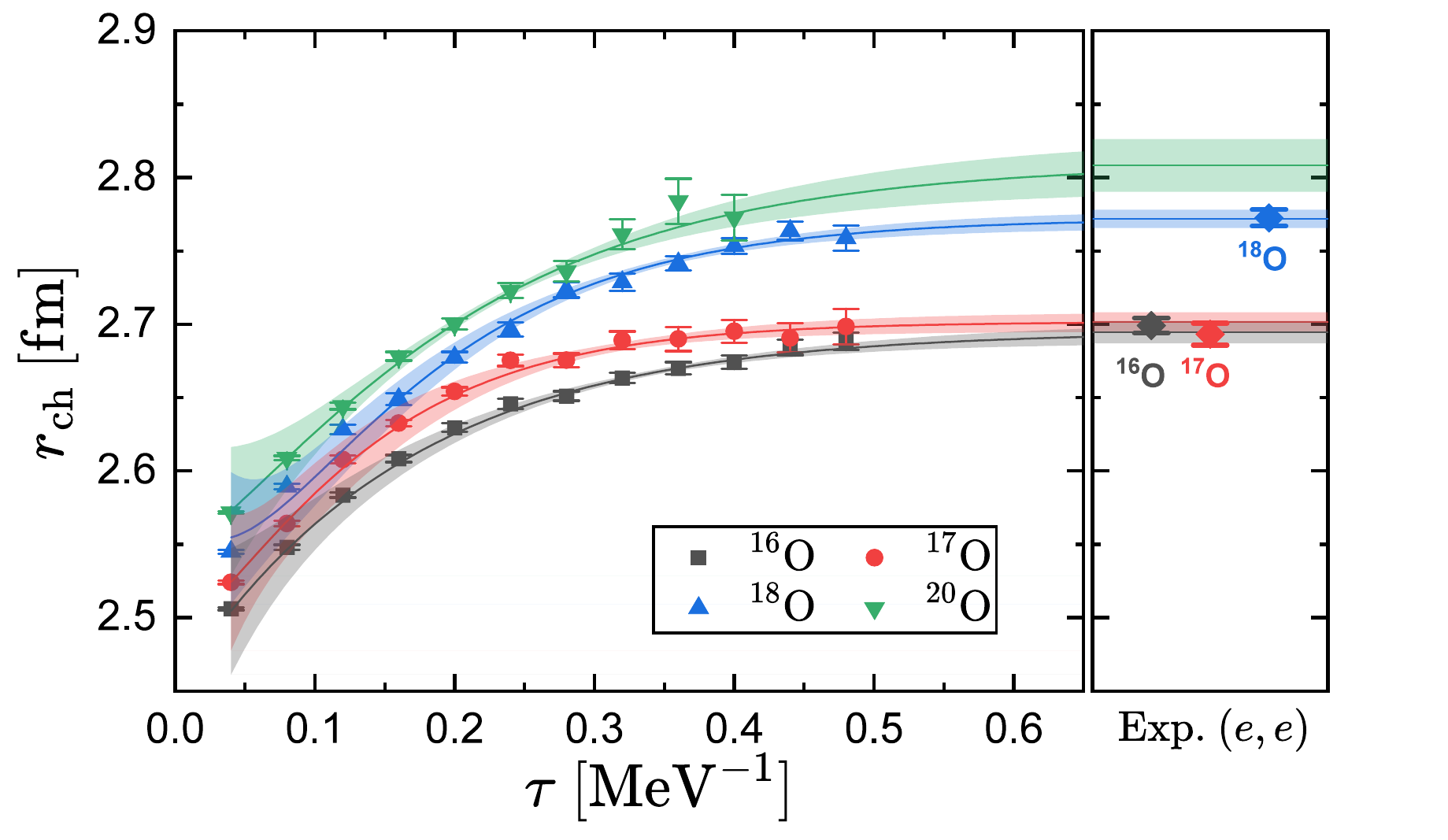}\\
  \caption{Left: Charge radii of oxygen isotopes calculated using the partial pinhole algorithm with the N$^3$LO interaction as functions of the projection time $\tau$.
  The black squares, red circles, blue upward triangles, and green downward triangles represent the results for $^{16}$O, $^{17}$O, $^{18}$O, and $^{20}$O, respectively.
  The solid lines denote fits using a double-exponential function, with the shaded bands indicating the 1$\sigma$ uncertainties.
  Right: Diamonds indicate the charge radii deduced from electron scattering data for $^{16}$O (black), $^{17}$O (red), and $^{18}$O (blue). Horizontal lines represent the extrapolated values at $\tau \to \infty$ from the left panel, with shaded bands showing the $1\sigma$ uncertainties.
  \label{fig1}}
\end{figure}

\begin{figure}[!htbp]
  \centering
  \includegraphics[width=0.45\textwidth]{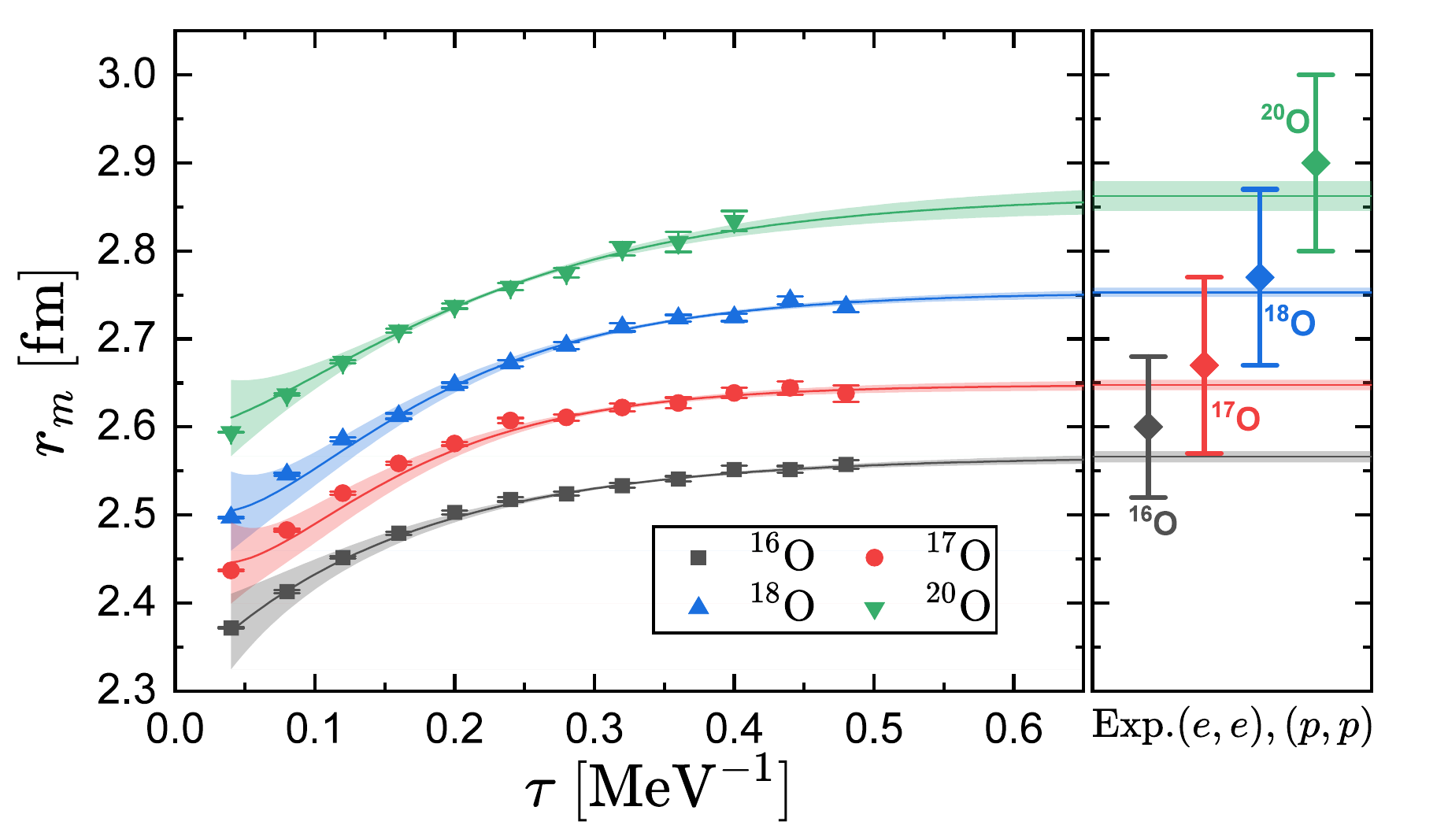}\\
  \caption{Left: Matter radii of oxygen isotopes calculated using the partial pinhole algorithm with the N$^3$LO interaction as functions of the projection time $\tau$.
  The black squares, red circles, blue upward triangles, and green downward triangles represent the results for $^{16}$O, $^{17}$O, $^{18}$O, and $^{20}$O, respectively.
  The solid lines denote fits using a double-exponential function, with the shaded bands indicating the 1$\sigma$ uncertainties.
  Right: Diamonds indicate the matter radii deduced from electron scattering and proton scattering data for $^{16}$O (black), $^{17}$O (red), $^{18}$O (blue), and $^{20}$O (green). Horizontal lines represent the extrapolated values at $\tau \to \infty$ from the left panel, with shaded bands showing the $1\sigma$ uncertainties.
  \label{fig2}}
  \end{figure}

\begin{table}[!htbp]
  \centering
  \caption{\label{Tab1} The charge and matter radii (units are fm) calculated by the partial pinhole algorithm using the N$^3$LO interaction, where the theoretical uncertainties are estimated using ten sets of three-nucleon forces. 
  The available experiment data are also given for comparison. 
  Exp. $(e,e)$~\cite{Angeli:2013epw}: charge radii deduced from electron scattering data.
  Exp. $(e,e)$, $(p,p)$~\cite{Lapoux:2016exf}: matter radii deduced from electron scattering and proton scattering data.
  Exp. $\sigma_{I}$~\cite{Ozawa:2000gx}: matter radii deduced from interaction cross sections.
  Exp. $\sigma_{cc}$~\cite{Kaur:2022yoh}: matter radii deduced from charge changing cross sections.
  }
\begin{ruledtabular}
\begin{tabular}{ccccc}
                         &$^{16}$O         &$^{17}$O      &$^{18}$O       &$^{20}$O\\
\hline
  $r_{\rm ch}$ [NLEFT]   &$2.704(17)$      &$2.709(15)$   &$2.768(17)$    &$2.810(32)$ \\
  Exp. $(e,e)$           &$2.699(5)$       &$2.693(8)$    &$2.776(2)$     &\\
  \hline
  $r_m$ [NLEFT]           &$2.576(17)$     &$2.651(14)$   &$2.744(19)$    &$2.863(33)$ \\
  Exp. $(e,e)$, $(p,p)$   &$2.60(8)$       &$2.67(10)$    &$2.77(10)$     &$2.90(10)$\\
  Exp. $\sigma_{I}$       &$2.54(2)$       &$2.59(5)$     &$2.61(8)$      &$2.69(3)$\\
  Exp. $\sigma_{cc}$      &$2.57(2)$       &              &$2.64(8)$      &$2.71(3)$\\
\end{tabular}
\end{ruledtabular}
\end{table}

Figure~\ref{fig1} shows the charge radii of $^{16}$O, $^{17}$O, $^{18}$O, and $^{20}$O calculated using the partial pinhole algorithm together with the chiral interactions at N$^3$LO, where a specified set of three-body forces is adopted~\cite{SM}.
Due to the substantial suppression of the sign oscillations, we are now able to reach significantly larger imaginary times as compared to
earlier works. For $^{16}$O, $^{17}$O, and $^{18}$O, the calculated radii exhibit a clear exponential decay behavior.
A double-exponential fit is used to account for residual excited-state contamination and extrapolate the results to $\tau \to \infty$; see the Supplemental Material~\cite{SM} for details.
The uncertainties of the extrapolation are quantified using the covariance matrix obtained from the nonlinear fit.
The extrapolated charge radii at $\tau\to\infty$ are $2.695(8)$, $2.702(7)$, and $2.772(6)$~fm, respectively.
Compared to the corresponding results presented in Ref.~\cite{Elhatisari:2022zrb}, the 1$\sigma$ uncertainties are significantly reduced.
This substantial improvement in precision enables a detailed investigation of the subtle isotopic trend in the charge radii of oxygen isotopes.
This level of precision also makes it feasible to probe smaller effects such as relativistic spin-orbit corrections and two-nucleon currents in the future.
For $^{20}$O, however, the stronger sign problem arising from its larger proton–neutron asymmetry prevents the simulation from reaching very large projection times, and the extrapolation is terminated at $\tau = 0.4~{\rm MeV}^{-1}$.
As a result, the extrapolated charge radius exhibits a larger uncertainty.

The extrapolated charge radii are summarized in Table~\ref{Tab1}, where results using nine additional sets of three-nucleon forces are included to estimate theoretical uncertainties~\cite{Elhatisari:pre}. 
Note that the central values and the corresponding errors are shifted compared to the extrapolated ones in Fig.~\ref{fig1}. 
The results show that $^{16}$O and $^{17}$O have very similar charge radii, while $^{18}$O exhibits a noticeable increase, consistent with experimental data~\cite{Angeli:2013epw}.
We note that, although $^{16}$O is commonly described as a spherical doubly magic nucleus within shell-model-based approaches, NLEFT simulations predict a distinct tetrahedral arrangement of alpha clusters~\cite{Epelbaum:2013paa}.
Similar alpha-cluster structures in $^{8}$Be~\cite{Shen:2024qzi}, $^{12}$C~\cite{Shen:2022bak}, and $^{20}$Ne~\cite{Harris:2025zsh} have also been predicted by NLEFT in their ground states.
Supporting this picture, previous NLEFT studies~\cite{Shen:2022bak} demonstrated that the ground state of $^{12}$C robustly converges to an equilateral triangular arrangement of alpha clusters independent of the choice of the initial state.
By analogy, assuming that $^{16}$O similarly forms a stable tetrahedral cluster configuration as indicated by NLEFT calculations, this intrinsic clustering could explain the success of NLEFT in reproducing oxygen isotope charge radii.

We also compute the matter radii of oxygen isotopes using the partial pinhole algorithm.
The results for $^{16}$O, $^{17}$O, $^{18}$O, and $^{20}$O are shown in Fig.~\ref{fig2}.
Similar to the charge radii shown in Fig.~\ref{fig1}, we perform a double-exponential fit to extrapolate the data to $\tau \to\infty$; see~\cite{SM} for details.
In contrast to the charge radii, the matter radii increase monotonically with mass number.
The extrapolated matter radii are listed in Table~\ref{Tab1}, along with values obtained using nine additional sets of three-nucleon forces~\cite{Elhatisari:pre}.

In Table~\ref{Tab1}, we also compare our results with experimental matter radii extracted from electron and proton scattering [$(e,e)$ and $(p,p)$]~\cite{Lapoux:2016exf}, interaction cross sections ($\sigma_I$)~\cite{Ozawa:2000gx}, and charge-exchange cross sections ($\sigma_{cc}$)~\cite{Kaur:2022yoh}.
Unlike charge radii, the extraction of matter radii often involves some degree of model dependence, leading to discrepancies among different experimental methods.
For example, matter radii derived from low-energy proton elastic scattering tend to be systematically larger than those extracted from $\sigma_I$ and $\sigma_{cc}$ measurements.
The most notable case is $^{20}$O, where differences up to 0.2~fm are observed among the three different measurements.
As discussed in Ref.~\cite{Lapoux:2016exf}, the $(e,e)$ and $(p,p)$ data provide a more reliable descriptions, since the correlations in the target are usually not included in the other methods.
From Table~\ref{Tab1}, we can find our calculated matter radii agree well with the values extracted from $(e,e)$ and $(p,p)$ data,
and thus support the statement given in Ref.~\cite{Lapoux:2016exf}.

\section{Summary and perspectives}
We have performed an {\em ab initio} calculation of the charge and matter radii of oxygen isotopes from $^{16}$O to $^{20}$O using nuclear lattice effective field theory (NLEFT) with the  high-fidelity N$^3$LO chiral interactions.
To mitigate the Monte Carlo sign problem in nuclear radius calculations, we introduce the {\em partial pinhole algorithm}, which reduces the rank of the many-body density operator in the standard pinhole approach and significantly lowers the statistical uncertainties.
This method enables access to much larger projection times and extends the applicability to more neutron-rich and proton-rich nuclei.
Our results accurately reproduce the experimental charge radii of $^{16}$O, $^{17}$O, and $^{18}$O, demonstrating the capability of NLEFT in describing nuclear structure properties.
The charge radius of $^{20}$O is predicted to be $2.810(32)$~fm.
The calculated matter radii show excellent agreement with values extracted from low-energy electron and proton scattering, while deviating from those inferred from interaction and charge-exchange cross sections.
These results underscore the need for further experimental clarification, particularly for neutron-rich isotopes, with improved precision and reduced model dependence.

As a general and versatile method, the partial pinhole algorithm holds broad potential future applications.
It can be integrated with other Monte Carlo techniques, such as the perturbative quantum Monte-Carlo method~\cite{Lu:2021tab},
and has already been successfully applied to the calculation of additional observables, including electromagnetic transitions~\cite{Shen:2024qzi}.
Its ability to suppress sign oscillations while maintaining high accuracy suggests wide applicability not only across nuclear systems, but also to broader classes of quantum many-body problems.

\begin{acknowledgments}
We are grateful for discussions with
Fabian Hildenbrand, Timo L\"ahde, Dean Lee, Bing-Nan Lu, Yuan-Zhuo Ma, Xiang-Xiang Sun, and Shuang Zhang.
This work was supported in part by the European
Research Council (ERC) under the European Union's Horizon 2020 research
and innovation programme (grant agreement No. 101018170),
and by the CAS President's International Fellowship Initiative (PIFI) (Grant No.~2025PD0022). The work of SE is supported
in part by the Scientific and Technological Research Council of Turkey (TUBITAK project no. 123F464). The authors gratefully acknowledge the Gauss Centre for Supercomputing e.V. (www.gauss-centre.eu)
for funding this project by providing computing time on the GCS Supercomputer JUWELS
at J\"ulich Supercomputing Centre (JSC). Furthermore, the authors gratefully acknowledge the computing time provided on the high-performance computer
HoreKa by the National High-Performance Computing Center at KIT (NHR@KIT). This center is
jointly supported by the Federal Ministry of Education and Research and the Ministry of Science,
Research and the Arts of Baden-Württemberg, as part of the National High-Performance Computing
(NHR) joint funding program (https://www.nhr-verein.de/en/our-partners). HoreKa is partly funded
by the German Research Foundation (DFG).
\end{acknowledgments}

%references for main text

\clearpage

%%%%%%%%%%%%%%%%%%%%%%%%%%%%%%%%%%%%%%%%%%%%%%%%%%%%%%%%%%%%%%%%%%%%%
%Supplemental Materials
%%%%%%%%%%%%%%%%%%%%%%%%%%%%%%%%%%%%%%%%%%%%%%%%%%%%%%%%%%%%%%%%%%%%%

\beginsupplement
\onecolumngrid

\section{Supplemental Materials}
\subsection{Lattice Hamiltonian and wave function matching with the N$^3$LO interaction}
In our lattice calculations, we combine the partial pinhole algorithm with the wave function matching method introduced in Ref.~\cite{Elhatisari:2022zrb_sm}.
In this approach, the original chiral  Hamiltonian is unitarily transformed into a new high-fidelity Hamiltonian $H$, whose wave function matches those of a computationally simpler Hamiltonian $H_S$ at short distances, up to a specified matching radius.
This construction ensures rapid convergence of the perturbative expansion in $H - H_S$.

The full formalism of the wave function matching method is given in Ref.~\cite{Elhatisari:2022zrb_sm}, and here we summarize the key elements relevant to our study.

The simplified Hamiltonian $H_S$ is constructed from a leading-order chiral effective field theory ($\chi$EFT) interaction,
\begin{equation}\label{eq:H_S}
  H_S = K + \frac{1}{2}c_{\rm SU(4)}\sum_{\bm{n}}:\tilde{\rho}^2(\bm{n}): + V_{\rm OPE}^{\Lambda_\pi=180},
\end{equation}
where $K$ denotes the kinetic energy term  with nucleon mass $m = 938.92$ MeV, and the colons $::$ indicate normal ordering.
The smeared density operator $\tilde{\rho}(\bm{n})$ includes both local and nonlocal components and is defined as
\begin{equation}
   \tilde{\rho}(\bm{n}) = \sum_{\sigma\tau}\tilde{a}^\dagger_{\sigma\tau}(\bm{n})\tilde{a}^{}_{\sigma\tau}(\bm{n}) + s_L\sum_{|\bm{n}'-\bm{n}|=1}\sum_{\sigma\tau}\tilde{a}^\dagger_{\sigma\tau}(\bm{n}')\tilde{a}^{}_{\sigma\tau}(\bm{n}'),
\end{equation}
with smeared creation and annihilation operators given by
\begin{equation}\label{eq:tildea_sm}
  \tilde{a}^{(\dagger)}_{\sigma\tau}(\bm{n}) = a^{(\dagger)}_{\sigma\tau}(\bm{n}) + s_{NL}\sum_{|\bm{n}'-\bm{n}|=1}a^{(\dagger)}_{\sigma\tau}(\bm{n}').
\end{equation}
Throughout our calculations, we use a local smearing parameter $s_L = 0.07$ and a nonlocal smearing parameter $s_{NL} = 0.5$.
The term $V_{\rm OPE}^{\Lambda_\pi=180}$ denotes the one-pion-exchange (OPE) potential at leading order, regulated with a cutoff $\Lambda_\pi = 180$~MeV.

The high-fidelity $\chi$EFT Hamiltonian $H$ at N$^3$LO is given by
\begin{align}\label{eq:H_N3LO}
H = K + V_{\rm OPE}^{\Lambda_\pi=300} + V_{\rm Coulomb} + V_{\rm 3N}^{Q^3} + V_{\rm 2N}^{Q^4} + W_{\rm 2N}^{Q^4} + V_{\rm 2N,WFM}^{Q^4} + W_{\rm 2N,WFM}^{Q^4}.
\end{align}
Here, the OPE potentials $V_{\rm OPE}^{\Lambda_\pi=300}$ has a larger cutoff $\Lambda_\pi=300$~MeV compared to $H_S$.
The full Hamiltonian also includes the Coulomb interaction ($V_{\rm Coulomb}$), the three-nucleon (3N) interaction at order $Q^3$ ($V_{\rm 3N}^{Q^3}$), two-nucleon (2N) short-range interactions at order $Q^4$ ($V_{\rm 2N}^{Q^4}$),
Galilean-invariance-restoring (GIR) terms ($W_{\rm 2N}^{Q^4}$), and wave function-matching interaction ($V_{\rm 2N,WFM}^{Q^4}$ as well as the GIR correction of the wave function matching interaction $W^{Q^4}_{\rm 2N,WFM}$. The contribution of $V_{\rm 2N,WFM}^{Q^4}$ to nuclear radii is found to be negligible, while its inclusion significantly increases computational cost. Therefore, this term is omitted in our simulations.

The 3N interactions at N$^{2}$LO consists of a contact potential, one-pion exchange potential, and two-pion exchange potential.
Same as Ref.~\cite{Elhatisari:2022zrb_sm}, the two-pion exchange potential is defined in a standard way.
The three-nucleon contact and three-nucleon one-pion exchange potentials can be further classified local and nonlocal potentials.
The general forms for local potentials read,
\begin{align}
   [V^{(d)}_{c_E}]_{abc} &= [c^{(d)}_{E}]_{abc}\sum_{\bm{n},\bm{n}',\bm{n}''}\rho^{(d)}(\bm{n}) \rho^{(d)}(\bm{n}') \rho^{(d)}(\bm{n}'')\delta_{|\bm{n}-\bm{n}'|^2,a}\delta_{|\bm{n}-\bm{n}''|^2,b}\delta_{|\bm{n}'-\bm{n}''|^2,c},\\
   [V_{c_E,s_L}]_{abc} &= [c_{E,s_L}]_{abc}\sum_{\bm{n},\bm{n}',\bm{n}''}\rho^{(1)}(\bm{n}) \rho^{(1)}(\bm{n}') \rho^{(0)}(\bm{n}'')\delta_{|\bm{n}-\bm{n}'|^2,a}\delta_{|\bm{n}-\bm{n}''|^2,b}\delta_{|\bm{n}'-\bm{n}''|^2,c},\\
   V_{c_D}^{(d)} &= -\frac{c_D^{(d)}g_A}{4F_\pi^4\Lambda_\chi}\sum_{\bm{n},S,I} \sum_{\bm{n}',S'} :\rho^{(0)}_{S',I}(\bm{n}')f_{S',S}(\bm{n}'-\bm{n})\rho^{(d)}_{S,I}(\bm{n})\rho^{(d)}(\bm{n}):,
\end{align}
with $f_{S',S}$ the locally-regulated pion correlation function,
\begin{equation}
   f_{S',S}(\bm{n}'-\bm{n}) = \frac{1}{L^3}\sum_{\bm{q}}\frac{q_Sq_{S'}e^{-\bm{q}\cdot(\bm{n}'-\bm{n})-(\bm{q}^2+M_\pi^2)/\Lambda_\pi^2}}{\bm{q}^2+M_\pi^2},
\end{equation}
and the ones for nonlocal potentials read,
\begin{align}
%   [ V_{c_E}^{(d)} ]^{s_{NL}}_{abc} &= [c_{E}^{(d)}]^{s_{NL}}_{abc}
%\sum_{\bm{n},\bm{n}',\bm{n}''}\tilde{\rho}^{(d)}(\bm{n})\tilde{\rho}^{(d)}(\bm{n}')\tilde{\rho}^{(d)}(\bm{n}'')
%\delta_{|\bm{n}-\bm{n}'|^2,a}\delta_{|\bm{n}-\bm{n}''|^2,b}\delta_{|\bm{n}'-\bm{n}''|^2,c},\\
   V_{c_E,s_{NL}}^{(d)}  &= c_{E,s_{NL}}^{(d)}
\sum_{\bm{n}}\tilde{\rho}^{(d)}(\bm{n})\tilde{\rho}^{(d)}(\bm{n})\tilde{\rho}^{(d)}(\bm{n}),\\
V_{c_D, s_{NL}}^{(d)} &= -\frac{c_{D,s_{NL}}^{(d)}g_A}{4 F_\pi^4 \Lambda_\chi}\sum_{\bm{n},S,I} \sum_{\bm{n}',S'}:\rho^{(0)}_{S',I}(\bm{n}')f_{S',S}(\bm{n}' - \bm{n})\tilde{\rho}^{(d)}_{S,I}(\bm{n})\tilde{\rho}^{(d)}(\bm{n}):.
\end{align}
Here, the locally smeared density operators are defined as,
\begin{align}
  \rho^{(d)}(\bm{n}) &= \sum_{\sigma\tau}a^\dagger_{\sigma\tau}(\bm{n})a_{\sigma\tau}(\bm{n}) + s_L\sum_{|\bm{n}'-\bm{n}|^2=1}^{d} \sum_{\sigma\tau}a^\dagger_{\sigma\tau}(\bm{n}')a^{}_{\sigma\tau}(\bm{n}'),\\
  \rho^{(d)}_{S,I}(\bm{n}) &=\sum_{\sigma\tau\sigma'\tau'}a_{\sigma\tau}^\dagger(\bm{n})[\sigma_S]_{\sigma\sigma'}[\sigma_I]_{\tau\tau'}a_{\sigma'\tau'}(\bm{n}) + s_L\sum_{|\bm{n}'-\bm{n}|^2=1}^{d}\sum_{\sigma\tau\sigma'\tau'}a^\dagger_{\sigma\tau}(\bm{n}')[\sigma_S]_{\sigma\sigma'}[\sigma_I]_{\tau\tau'}a^{}_{\sigma'\tau'}(\bm{n}'),
\end{align}
and the non-locally smeared density operators are defined as,
\begin{align}
  \tilde{\rho}^{(d)}(\bm{n}) &= \sum_{\sigma\tau}\tilde{a}^\dagger_{\sigma\tau}(\bm{n})\tilde{a}_{\sigma\tau}(\bm{n}) + s_L\sum_{|\bm{n}'-\bm{n}|^2=1}^{d} \sum_{\sigma\tau}\tilde{a}^\dagger_{\sigma\tau}(\bm{n}')\tilde{a}^{}_{\sigma\tau}(\bm{n}'),\\
  \tilde{\rho}^{(d)}_{S,I}(\bm{n}) &=\sum_{\sigma\tau\sigma'\tau'}\tilde{a}_{\sigma\tau}^\dagger(\bm{n})[\sigma_S]_{\sigma\sigma'}[\sigma_I]_{\tau\tau'}\tilde{a}_{\sigma'\tau'}(\bm{n}) + s_L\sum_{|\bm{n}'-\bm{n}|^2=1}^{d}\sum_{\sigma\tau\sigma'\tau'}\tilde{a}^\dagger_{\sigma\tau}(\bm{n}')[\sigma_S]_{\sigma\sigma'}[\sigma_I]_{\tau\tau'}\tilde{a}^{}_{\sigma'\tau'}(\bm{n}'),
\end{align}
where the smeared creation and annihilation operators are defined in Eq.~\eqref{eq:tildea_sm}. 
Here, the local smearing parameter $s_L$ is taken same as that in the simplified Hamiltonian $H_S$, i.e., $s_L=0.1$, while the  nonlocal smearing parameter $s_{NL}$ is taken as $0.1$, $0.2$, or $0.3$, which will be denoted as $s_{NL1}$, $s_{NL2}$, and $s_{NL3}$, respectively. 

\begin{figure}[!htbp]
  \centering
  \includegraphics[width=0.85\textwidth]{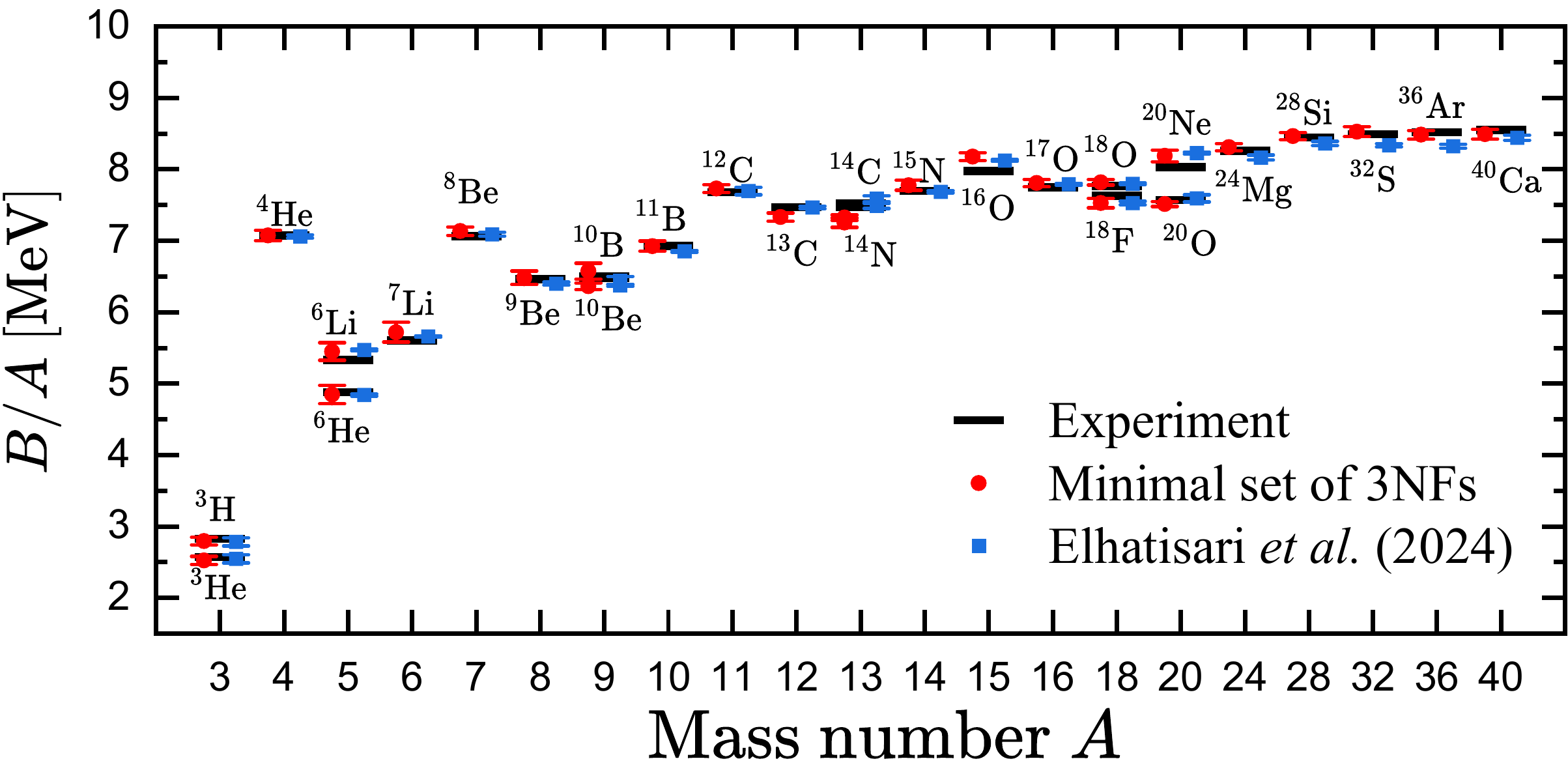}\\
  \caption{Nuclear binding energies $B$ from NLEFT in comparison to experimental data. The red circles summarise the results with ten minimal sets of 3NFs, and blue squares represent binding energies from Ref.~\cite{Elhatisari:2022zrb_sm}. The error bars show standard deviations.
  \label{fig1_SM}}
\end{figure}

All simulations are performed in periodic cubic lattices with a spatial lattice spacing of $a = 1.32$~fm. The low-energy constants (LECs) of the two-nucleon interaction are taken from Ref.~\cite{Elhatisari:2022zrb_sm}. 
For the 3N potential, we adopt a minimal set of three-body force terms (3NFs) that reduce the number of free LECs from eight to six while preserving the overall description of nuclear binding energies~\cite{Elhatisari:pre_SM}.
The information about the adopted ten sets of 3NFs are listed in Table.~\ref{tab:3NFs}, where the set with No. 1 is adopted for the Figs. 1 and 2 in main text. 
The binding energies with the comparison to experimental data  are shown in Fig.~\ref{fig1_SM} and Table \ref{tab:Be}, where ten sets of 3NFs have been used to estimate theoretical uncertainties.
The resulting average root mean square deviation (RMSD) is 0.101 MeV (1.4\%) for the average binding energy per nucleon, which is quantitatively comparable to the value 0.093 MeV (1.3\%) from Ref.~[21].

\begingroup
\renewcommand{\arraystretch}{1.4}
\begin{table}[!htbp]
  \centering
  \caption{\label{tab:3NFs} Adopted ten sets of three-nucleon force terms and the corresponding LECs.}
\begin{ruledtabular}
\begin{tabular}{lll}
        No.   & $V_{3N}$'s  & LECs $c$ [in lattice units] \\
   \hline
       1       &$V_{c_E,s_{NL1}}^{(0)}$, $V_{c_D}^{(3)}$,  $[V_{c_E}^{(0)}]_{044}$,  $[V_{c_E}^{(0)}]_{224}$,  $[V_{c_E}^{(0)}]_{125}$,  $[V_{c_E}^{(0)}]_{145}$
       &$0.561536$,$-1.148095$,$0.407892$, $0.193516$,$-0.227340$,$-0.022996$ \\
       2      &$[V_{c_E}^{(1)}]_{000}$, $V_{c_D}^{(3)}$, $[V_{c_E}^{(0)}]_{044}$, $[V_{c_E}^{(0)}]_{222}$, $[V_{c_E}^{(0)}]_{114}$, $[V_{c_E,s_L}]_{033}$ & $1.683416$, $-0.386143$, $0.852459$, $0.077566$, $-0.507509$, $-0.526746$ \\
  3 & $V_{c_E,s_{NL1}}^{(0)}$, $V_{c_D}^{(3)}$, $[V_{c_E}^{(0)}]_{044}$, $[V_{c_E}^{(0)}]_{224}$, $[V_{c_E}^{(0)}]_{125}$, $[V_{c_E}^{(0)}]_{235}$ & 0.562797, $-1.152401$, 0.332151, 0.209908, $-0.228472$, $-0.015002$ \\
  4 & $V_{c_D}^{(0)}$, $V_{c_D}^{(2)}$, $[V_{c_E}^{(0)}]_{044}$, $[V_{c_E}^{(0)}]_{222}$, $[V_{c_E}^{(0)}]_{224}$, $[V_{c_E}^{(0)}]_{125}$ & 3.219339, $-1.359388$, 0.284671, 0.006149, 0.167456, $-0.227336$ \\
  5 & $V_{c_D}^{(1)}$, $V_{c_D}^{(3)}$, $[V_{c_E}^{(0)}]_{033}$, $[V_{c_E}^{(0)}]_{044}$, $[V_{c_E}^{(0)}]_{222}$, $[V_{c_E}^{(0)}]_{125}$ & 1.983634, $-1.136031$, $-0.860450$, 1.147171, 0.078412, $-0.170220$ \\
  6 & $V_{c_E,s_{NL1}}^{(0)}$, $V_{c_D}^{(0)}$, $V_{c_D}^{(3)}$, $[V_{c_E}^{(0)}]_{044}$, $[V_{c_E}^{(0)}]_{224}$, $[V_{c_E}^{(0)}]_{125}$ & 0.205880, 1.642363, $-1.004686$, 0.328845, 0.179790, $-0.228399$ \\
  7 & $V_{c_D}^{(1)}$, $V_{c_D}^{(3)}$, $V_{c_D,s_{NL2}}^{(0)}$, $[V_{c_E}^{(0)}]_{044}$, $[V_{c_E}^{(0)}]_{224}$, $[V_{c_E}^{(0)}]_{125}$ & 2.084608, $-1.138905$, 0.002480, 0.306560, 0.173300, $-0.223509$ \\
  8 & $V_{c_D}^{(0)}$, $V_{c_D}^{(2)}$, $[V_{c_E}^{(0)}]_{044}$, $[V_{c_E}^{(0)}]_{224}$, $[V_{c_E}^{(0)}]_{125}$, $[V_{c_E}^{(0)}]_{145}$ & 3.200088, $-1.363258$, 0.332144, 0.193416, $-0.228399$, $-0.019028$ \\
  9 & [$V_{c_E}^{(0)}]_{000}$, $V_{c_D}^{(3)}$, $[V_{c_E}^{(0)}]_{224}$, $[V_{c_E}^{(0)}]_{125}$, $[V_{c_E,s_L}]_{044}$, $[V_{c_E}^{(0)}]_{334}$ & 2.359915, $-0.485001$, 0.164774, $-0.228399$, 0.169945, 0.002113 \\
 10 &[$V_{c_E}^{(0)}]_{000}$, $V_{c_E,s_{NL3}}^{(0)}$, $V_{c_D}^{(1)}$, $V_{c_D,s_{NL3}}^{(0)}$, $[V_{c_E}^{(0)}]_{224}$, $[V_{c_E}^{(0)}]_{125}$ & 4.315558, 0.039782, $-2.101720$, $-0.109547$, 0.145754, $-0.169632$ \\
\end{tabular}
\end{ruledtabular}
\end{table}
\endgroup

\begingroup
\renewcommand{\arraystretch}{1.2}
\begin{table}[!htbp]
  \centering
  \caption{\label{tab:Be} Nuclear binding energies $B$ (in MeV) from NLEFT in comparison to experimental data.}
\begin{ruledtabular}
\begin{tabular}{cccc}
       Nuclei   & Experiment  &NLEFT: Minimal set of 3NFs  &NLEFT: Elhatisari \textit{et al.} (2024)~\cite{Elhatisari:2022zrb_sm}\\
   \hline
       $^{3}$H     &$8.48$    &$8.39(0.16)$      &$8.35(0.17)$ \\
       $^{3}$He   &$7.72$    &$7.57(0.17)$      &$7.64(0.17)$ \\
       $^{4}$He   &$28.30$    &$28.30(0.28)$      &$28.24(0.09)$ \\
       $^{6}$He   &$29.27$    &$29.08(0.78)$      &$29.04(0.09)$ \\
       $^{6}$Li    &$31.99$    &$32.70(0.74)$      &$32.82(0.09)$ \\
       $^{7}$Li    &$39.24$    &$40.03(0.96)$      &$39.61(0.08)$ \\
       $^{8}$Be   &$56.50$    &$57.09(0.46)$      &$56.73(0.21)$ \\
       $^{9}$Be   &$58.17$    &$58.34(0.87)$      &$57.59(0.20)$ \\
       $^{10}$Be  &$64.97$    &$63.63(0.46)$      &$63.72(0.14)$ \\
       $^{10}$B    &$64.75$    &$65.74(1.13)$      &$64.46(0.56)$ \\
       $^{11}$B     &$76.2$    &$76.18(0.79)$      &$75.38(0.13)$ \\
       $^{12}$C    &$92.16$    &$92.77(0.66)$      &$92.36(0.61)$ \\
       $^{13}$C    &$97.11$    &$95.32(0.76)$      &$97.07(0.26)$ \\
       $^{14}$C    &$105.28$    &$102.55(0.53)$     &$104.87(0.51)$ \\
       $^{14}$N   &$104.66$    &$101.56(0.92)$     &$106.25(0.55)$ \\
       $^{15}$N   &$115.49$    &$116.66(1.05)$     &$115.29(0.22)$ \\
       $^{16}$O   &$127.62$    &$130.84(0.87)$     &$129.99(0.19)$ \\
       $^{17}$O   &$131.76$    &$132.74(0.88)$     &$132.47(0.26)$ \\
       $^{18}$O   &$139.81$    &$140.73(0.67)$     &$140.37(0.21)$ \\
       $^{20}$O  &$151.37$    &$150.30(0.70)$     &$151.9(1.01)$ \\
       $^{17}$F  &$137.37$    &$135.51(1.21)$     &$135.52(0.44)$ \\
       $^{20}$Ne &$160.65$    &$163.78(1.63)$     &$164.57(0.32)$ \\
       $^{24}$Mg &$198.26$    &$199.48(1.19)$     &$195.96(0.85)$ \\
       $^{28}$Si  &$236.54$    &$237.02(1.39)$     &$234.16(0.85)$ \\
       $^{32}$S   &$271.78$    &$272.89(2.23)$     &$266.79(0.69)$ \\
       $^{36}$Ar &$306.72$    &$305.49(2.18)$     &$299.68(0.99)$ \\
       $^{40}$Ca &$342.05$    &$339.68(2.70)$     &$337.71(1.34)$ \\
\end{tabular}
\end{ruledtabular}
\end{table}
\endgroup

\subsection{Perturbative formalism for the partial pinhole algorithm}
To explain the perturbative formalism, we begin by expressing the two-body correlation functions $G_{\alpha\beta}$ in terms of the transfer matrix,
\begin{align}\label{eq:Gab_sm}
  G_{\alpha\beta} = \frac{A(A - 1)}{M(M - 1)}
   \frac{ \displaystyle\sum_{\vec{c}} \left[ \langle \Psi_0 |\hat{M}^{L_t/2} \rho_M(\vec{c}\,) \hat{M}^{L_t/2}|\Psi_0\rangle \sum_{i<j} \delta_{\tau_i\alpha} \delta_{\tau_j\beta} (\bm{n}_i - \bm{n}_j)^2 \right] }{ \displaystyle\sum_{\vec{c}} \langle \Psi_0 | \hat{M}^{L_t/2}\rho_M(\vec{c}\,) \hat{M}^{L_t/2}| \Psi_0 \rangle },
\end{align}
with the definition,
\begin{equation}
  \hat{M} = :\exp(-a_t H):.
\end{equation}
To formulate the perturbative expansion, we decompose the high-fidelity Hamiltonian $H$ in Eq.~\eqref{eq:H_N3LO} as
\begin{equation}
  H = H_S + (H - H_S),
\end{equation}
where the leading-order Hamiltonian $H_S$ is used as the nonperturbative part, while the difference $(H - H_S)$ is treated as a perturbation.
Accordingly, the transfer matrix can be expanded to first order as
\begin{align}
   \hat{M} = :\exp(-a_t H): = :\exp(-a_t H_S): - a_t :\exp(-a_t H_S)(H - H_S): \equiv \hat{M}^{(0)} + \hat{M}^{(1)}.
\end{align}

Keeping terms up to $\mathcal{O}(\hat{M}^{(1)})$, the correlation function becomes
\begin{equation}
    G_{\alpha\beta} = \frac{A(A - 1)}{M(M - 1)}\frac{\mathcal{M}_{\alpha\beta}^{(0)}+\mathcal{M}_{\alpha\beta}^{(1)}}{\mathcal{M}^{(0)}+\mathcal{M}^{(1)}}=\frac{A(A - 1)}{M(M - 1)}\left(\frac{\mathcal{M}_{\alpha\beta}^{(0)}}{\mathcal{M}^{(0)}} +\frac{\mathcal{M}_{\alpha\beta}^{(1)}}{\mathcal{M}^{(0)}} -\frac{\mathcal{M}_{\alpha\beta}^{(0)}\mathcal{M}^{(1)}}{(\mathcal{M}^{(0)})^2} \right),
\end{equation}
The nonperturbative amplitudes are defined as
\begin{subequations}
   \begin{align}
      &\mathcal{M}^{(0)} =  \sum_{\vec{c}} \langle \Psi_0 | (\hat{M}^{(0)})^{L_t/2}\rho_M(\vec{c}\,) (\hat{M}^{(0)})^{L_t/2}| \Psi_0 \rangle,\\
      &\mathcal{M}^{(0)}_{\alpha\beta} =  \sum_{\vec{c}}{\bigg [} \langle \Psi_0 | (\hat{M}^{(0)})^{L_t/2}\rho_M(\vec{c}\,) (\hat{M}^{(0)})^{L_t/2}| \Psi_0 \rangle\sum_{i<j} \delta_{\tau_i\alpha} \delta_{\tau_j\beta} (\bm{n}_i - \bm{n}_j)^2{\bigg ]}~.
   \end{align}
\end{subequations}
The perturbative amplitudes $\mathcal{M}^{(1)}$ and $\mathcal{M}_{\alpha\beta}^{(1)}$ are obtained by replacing one of the $\hat{M}^{(0)}$ factors with $\hat{M}^{(1)}$ at each time slice and summing over all possible insertion points,
\begin{subequations}
   \begin{align}
      &\mathcal{M}^{(1)} =  \sum_{\vec{c}}\sum_{n_t=0}^{L_t/2-1} \langle \Psi_0 | (\hat{M}^{(0)})^{L_t/2-n_t-1}M^{(1)}(\hat{M}^{(0)})^{n_t}\rho_M(\vec{c}\,) (\hat{M}^{(0)})^{L_t/2}| \Psi_0 \rangle+{\rm c.c.},\\
      &\mathcal{M}^{(1)}_{\alpha\beta} =  \sum_{\vec{c}}\sum_{n_t=0}^{L_t/2-1} {\bigg [}\langle \Psi | (\hat{M}^{(0)})^{L_t/2-n_t-1}M^{(1)}(\hat{M}^{(0)})^{n_t}\rho_M(\vec{c}\,)  (\hat{M}^{(0)})^{L_t/2}| \Psi \rangle\sum_{i<j} \delta_{\tau_i\alpha} \delta_{\tau_j\beta} (\bm{n}_i - \bm{n}_j)^2{\bigg ]} +{\rm c.c.}.
   \end{align}
\end{subequations}
In the calculation of all amplitudes $\mathcal{M}^{(0)}$, $\mathcal{M}^{(0)}_{\alpha\beta}$, $\mathcal{M}^{(1)}$, and $\mathcal{M}^{(1)}_{\alpha\beta}$, $ :\exp(-a_tH_S):$ will be used in the auxiliary-field formalism, and the other terms involving sophisticated $\chi$EFT interactions are evaluated via Jacobi formulas.

\subsection{The distance between two lattice sites}

\begin{figure}[!htbp]
  \centering
  \includegraphics[width=0.45\textwidth]{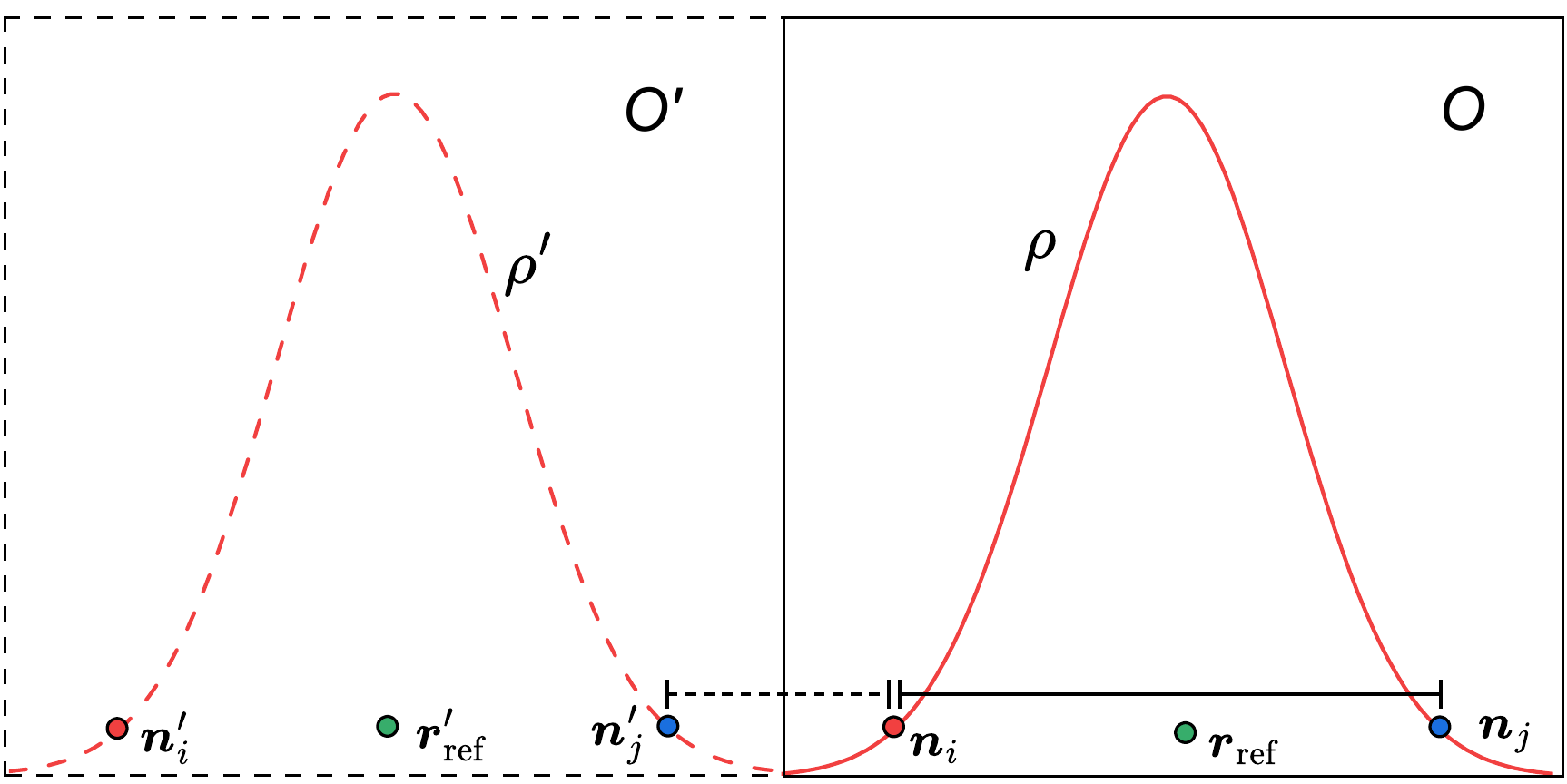}\\
  \caption{Schematic figure to show the additional finite volume effects introduced by the periodic boundary conditions.
  The frame $O'$ is a periodic copy of the frame $O$.
  \label{fig2_SM}}
  \end{figure}

To evaluate the correlation function $G_{\alpha\beta}$, one needs to calculate the squared distance between two lattice sites, $(\bm{n}_i-\bm{n}_j)^2$.
Due to the periodic boundary conditions used in our lattice calculations, the default manner is to minimize $(\bm{n}_i-\bm{n}_j)^2$  by considering all periodic images of the lattice sites. We denote the minimal-distance on the lattice as $(\widetilde{\bm{n}_i-\bm{n}_j})^2$.
However, this conventional method introduces additional finite volume effects.
This can be seen in Fig.~\ref{fig2_SM}, the physical distance between $\bm{n}_i$ and $\bm{n}_j$ will be replaced by the distance between $\bm{n}_i$ and $\bm{n}_j'$ due to the periodic boundary conditions.

Inspired by the fact that the nuclear radii are defined with respect to the center of mass, we calculate $(\bm{n}_i-\bm{n}_j)^2$ by introducing a reference point chosen to be close to the nuclear center of mass, denoted as $\bm{r}_{\rm ref}$.
From Fig.~\ref{fig2_SM}, we can find that the relative vector $(\widetilde{\bm{n}_i-\bm{r}_{\rm ref}})$ is always physically correct.
We further require that $\bm{r}_{\rm ref}$ should coincide exactly with the center of the periodic box so that it is symmetric with respect to all lattice directions.
Consequently, the coordinates of $\bm{r}_{\rm ref}$ depend on whether the box size $L$ is odd or even, e.g. along the $x$-direction,
\begin{equation}
   x_{\rm ref} =
   \begin{cases}
     n_x, &L={\rm odd},\\
     n_x +1/2, &L={\rm even},
   \end{cases}
\end{equation}
with $n_x$ an integer number.
The reference points $\bm{r}_{\rm ref}$ are then determined by minimizing
\begin{equation}
     \sum_{\bm{n}}\langle\rho(\bm{n})\rangle(\widetilde{\bm{n}-\bm{r}_{\rm ref}})^2,
\end{equation}
where the expectation of one-body density $\langle\rho(\bm{n})\rangle$ is defined as,
\begin{equation}
  \langle\rho(\bm{n})\rangle = \frac{\sum_{\sigma\tau}\langle\Psi|\rho_{\sigma\tau}(\bm{n}):\rho_{\sigma_i\tau_i}(\bm{n}_i)\rho_{\sigma_j\tau_j}(\bm{n}_j):|\Psi\rangle}{\langle\Psi|:\rho_{\sigma_i\tau_i}(\bm{n}_i)\rho_{\sigma_j\tau_j}(\bm{n}_j):|\Psi\rangle}.
\end{equation}
With the determined reference point, we first determine the relative vector of $\bm{n}_{i,j}$ with considering the periodicity property, and
$(\bm{n}_i-\bm{n}_j)^2$ is then evaluated as,
\begin{equation}
   [(\widetilde{\bm{n}_i-\bm{r}_{\rm ref}})-(\widetilde{\bm{n}_j-\bm{r}_{\rm ref}})]^2= (\widetilde{\bm{n}_i-\bm{r}_{\rm ref}})^2-2(\widetilde{\bm{n}_i-\bm{r}_{\rm ref}})\cdot(\widetilde{\bm{n}_j-\bm{r}_{\rm ref}}) + (\widetilde{\bm{n}_j-\bm{r}_{\rm ref}})^2.
\end{equation}

\subsection{Benchmark with the simple Hamiltonian $H_{S}$}
In this section, we benchmark the nuclear radii calculated using the partial pinhole algorithm against those obtained from the standard pinhole algorithm~\cite{Elhatisari:2017eno_SM}.
In the standard pinhole algorithm, the positions of $A$-nucleons, denoted as $\bm{n}_i$, are sampled according to the $A$-body density amplitudes.
With these $A$ pinhole coordinates, one can compute density correlations relative to the center of mass, and thus the nuclear radii can be obtained~\cite{Lu:2018bat_SM}.

For benchmarking purposes, we use the simplified Hamiltonian given in  Eq.~\eqref{eq:H_S}.
To facilitate perturbation theory, we split $H_S$ into a nonperturbative part $H^{(0)}$ and a perturbative correction $H^{(1)}$, given explicitly by,
\begin{align}
   &H^{(0)} = K + (1-\lambda)\cdot\frac{1}{2}c_{\rm SU(4)}\sum_{\bm{n}}:\tilde{\rho}^2(\bm{n}): + V_{\rm OPE}^{\Lambda_\pi=180},\\
   &H^{(1)} = \lambda\cdot\frac{1}{2}c_{\rm SU(4)}\sum_{\bm{n}}:\tilde{\rho}^2(\bm{n}):,
\end{align}
with parameter $\lambda=0.1$.
Such a value is chosen to ensures the validity of first-order perturbation theory.

\begin{figure}[!htbp]
  \centering
  \includegraphics[width=0.45\textwidth]{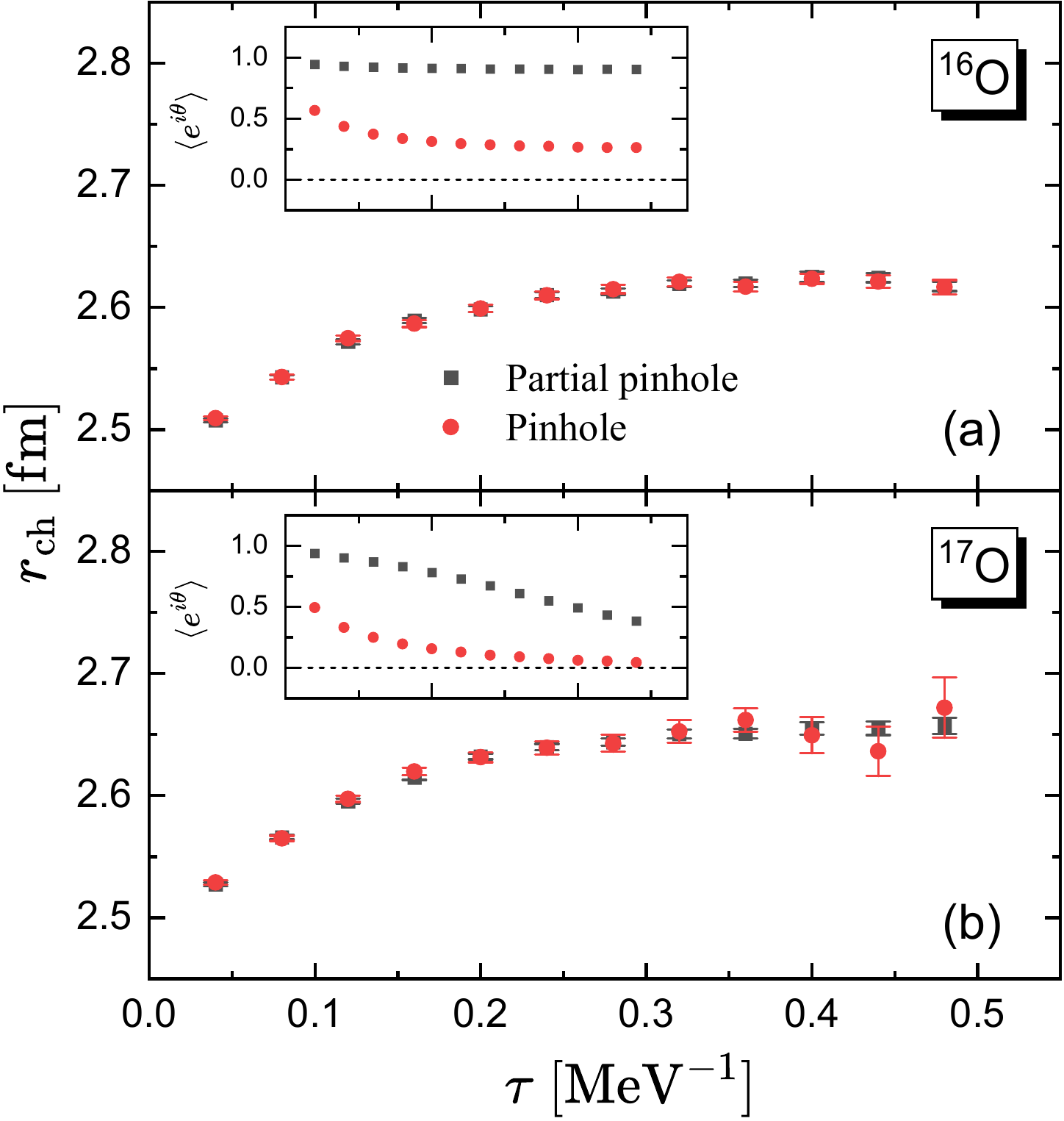}\\
  \caption{Charge radii of $^{16}$O (top) and $^{17}$O (bottom) calculated with simple Hamiltonian as function of the projection time $\tau$.
    The black squares and red circles represent the results from the partial pinhole and the pinhole algorithm, respectively.
    The insets show the average phase factor for the partial pinhole and the pinhole algorithms as functions of the projection time $\tau$.
  \label{fig3_SM}}
  \end{figure}
\begin{figure}[!htbp]
  \centering
  \includegraphics[width=0.45\textwidth]{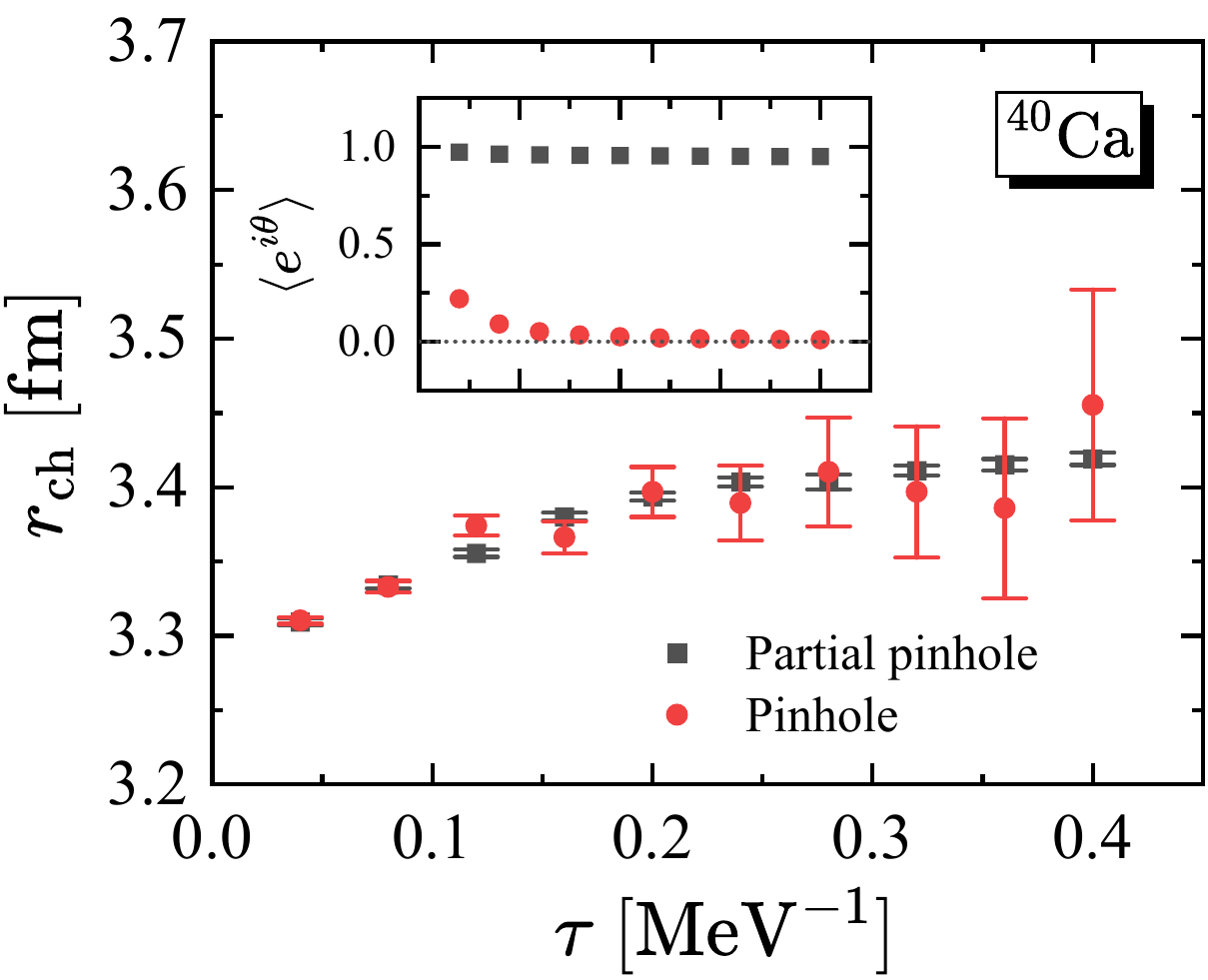}\\
  \caption{Same as Fig.~\ref{fig3_SM} but for $^{40}$Ca.
  \label{fig4_SM}}
  \end{figure}

As specific benchmarks, we compare the charge radii of $^{16}$O and $^{17}$O calculated using the partial pinhole and the standard pinhole algorithms in
Fig.~\ref{fig3_SM}.
We can find that both algorithms give us almost identical charge radii.
However,  the partial pinhole algorithm maintains a significantly large phase factor, especially noticeable for $^{17}$O at projection time longer than $0.4~{\rm MeV}^{-1}$.
As a result, the charge radii from partial pinhole algorithm are always shown with much smaller statistical uncertainties. For heavier nuclei, such as $^{40}$Ca, the difference becomes even more pronounced. As presented in Fig.~\ref{fig4_SM}, the average phase factor in the standard pinhole algorithm drops dramatically with the increase of $\tau$, a phase factor of less than $0.01$ at a projection time of $0.4~{\rm MeV}^{-1}$, while the partial pinhole algorithm achieves a phase factor as high as $0.95$.
As a results, the statistical uncertainties of the charge radius are greatly reduced by the partial pinhole algorithm. The advantage of the partial pinhole approach becomes even greater when employing the full N$^3$LO chiral interactions.

\begin{figure}[!htbp]
  \centering
  \includegraphics[width=0.45\textwidth]{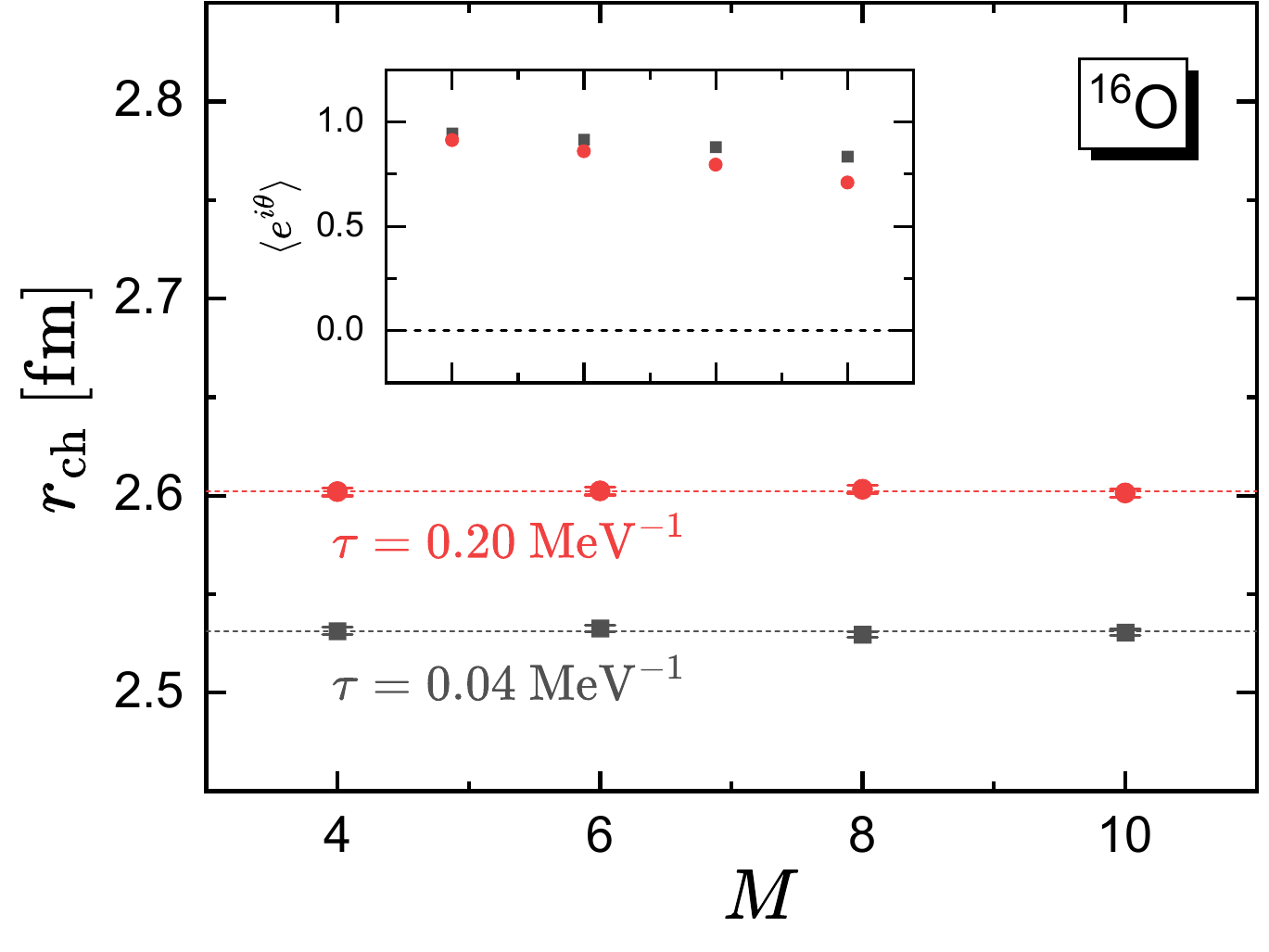}\\
  \caption{Charge radii calculated with simple Hamiltonian as functions of $M$ in $\rho_M$ of the partial pinhole algorithm.
    The black squares and red circles represent the results with the projection time $\tau=0.04~{\rm MeV}^{-1}$ and $0.20~{\rm MeV}^{-1}$, respectively.
    The inserts show the average phase factor as functions of $M$.
  \label{fig5_SM}}
  \end{figure}
To further illustrate the effectiveness of the partial pinhole algorithm, Fig.~\ref{fig5_SM} displays how charge radii depend on the number $M$ of nucleons included in the sampled density operator $\rho_M$, taking $^{16}$O as example. We can find that calculated charge radii are independent on the choice of $M$.
The average phase factors decrease gradually as $M$ increases, approaching the smaller values characteristic of the standard pinhole method in the limit $M = A$. This also explains how the partial pinhole algorithm mitigates the Monte Carlo sign problem encountered in nuclear radius calculation.

\subsection{Imaginary time extrapolation}
In Figs.~1 and 2 of the main text, we perform extrapolations to the $\tau \to \infty$ in order to extract the ground-state charge and matter radii.
For each observable $O^{(i)}$, we fit the calculated values at finite projection time $\tau$ using a double-exponential  ansatz,
\begin{equation}\label{eq:fit_eq}
O^{(i)} = O^{(i)}({\infty}) + O_1^{(i)}e^{-d\tau/2} + O_2^{(i)}e^{-d\tau},
\end{equation}
where $O^{(i)}({\infty})$, $O_1^{(i)}$, $O_2^{(i)}$, and the decay constant $d$ are fit parameters.
The observables considered in the fit include both charge and matter radii, computed at leading order and at N$^3$LO.
For each nucleus, all observables share the same decay parameter $d$ to ensure consistency in the treatment of excited-state contaminations.

The function form in Eq.~\eqref{eq:fit_eq} is motivated by the assumption that at sufficiently large imaginary time, and the wave function is dominated by the ground state and first excited state.
However, there is no strict a priori criterion for how large $\tau$ must be for this assumption to hold.
We therefore adopt a practical strategy in which we vary the lower bound $\tau_c$ of the fit imaginary-time window to identify a stable region.
Specifically, data points with $\tau < \tau_c$ are the suppressed by assigning them an artificially large uncertainty (e.g., 0.05~fm), thereby reducing their influence on the fit.
By scanning a range of $\tau_c$ and monitoring the stability of the extrapolated quantities, we identify a plateau region where the results become insensitive to $\tau_c$. This behavior signals a reliable window for extracting the ground-state properties.

The dependence of the extrapolated charge and matter radii on $\tau_c$ is presented in Tables~\ref{Tab1_SM}--\ref{Tab4_SM} for $^{16}$O, $^{17}$O, $^{18}$O, and $^{20}$O, respectively.
In all cases, we observe stable plateaus in the extrapolated charge and matter radii as $\tau_c$ increases.
Once the plateau is reached, the fitted parameter $O^{(i)}({\infty})$ in Eq.~\eqref{eq:fit_eq} exhibits a negligible variation.
Consequently, the fitted curves shown in Figs.~1 and 2 of the main text remain virtually unchanged.
This confirms the robustness of our extrapolated nuclear radii.
In the discussion of the main text, we adopt $\tau_c = 0.18~\mathrm{MeV}^{-1}$ for all extrapolations,
as it lies within the plateau region for the investigated oxygen isotopes and ensures stable and consistent extrapolation results.

\begin{table}[!htbp]
  \centering
  \caption{\label{Tab1_SM} Extrapolated charge radius $r_{\rm ch}$ and matter radius $r_m$ of $^{16}$O as functions of the lower bound $\tau_c$ of the imaginary-time window used in the double-exponential fit.}
\begin{ruledtabular}
\begin{tabular}{ccccccc}
  $\tau_c$~[MeV$^{-1}$]        & 0.02         &0.06         &0.10         &0.14          &0.18          &0.22\\
\hline
  $r_{\rm ch}$ [fm]            &$2.695(14)$   &$2.704(16)$  &$2.703(16)$  &$2.703(17)$   &$2.704(17)$   &$2.704(18)$\\
  $r_{m}$ [fm]                 &$2.569(15)$   &$2.576(16)$  &$2.575(16)$  &$2.575(17)$   &$2.576(17)$   &$2.576(18)$\\
\end{tabular}
\end{ruledtabular}
\end{table}

\begin{table}[!htbp]
  \centering
  \caption{\label{Tab2_SM} Same as Table~\ref{Tab1_SM} but for $^{17}$O.}
\begin{ruledtabular}
\begin{tabular}{ccccccc}
  $\tau_c$~[MeV$^{-1}$]        & 0.02         &0.06         &0.10         &0.14          &0.18         &0.22\\
\hline
  $r_{\rm ch}$ [fm]            &$2.711(14)$   &$2.711(15)$  &$2.711(15)$  &$2.711(15)$   &$2.709(15)$  &$2.708(16)$\\
  $r_{m}$ [fm]                 &$2.650(13)$   &$2.653(14)$  &$2.652(14)$  &$2.651(14)$   &$2.651(14)$  &$2.651(16)$\\
\end{tabular}
\end{ruledtabular}
\end{table}

\begin{table}[!htbp]
  \centering
  \caption{\label{Tab3_SM} Same as Table~\ref{Tab1_SM} but for $^{18}$O.}
\begin{ruledtabular}
\begin{tabular}{ccccccc}
  $\tau_c$~[MeV$^{-1}$]     &0.02          & 0.06        &0.10         &0.14        &0.18        &0.22\\
\hline
  $r_{\rm ch}$ [fm]         &$2.767(16)$   &$2.770(17)$  &$2.773(17)$  &$2.768(17)$ &$2.768(17)$ &$2.767(19)$\\
  $r_{m}$ [fm]              &$2.744(20)$   &$2.748(20)$  &$2.749(20)$  &$2.745(19)$ &$2.744(19)$ &$2.744(20)$ \\
\end{tabular}
\end{ruledtabular}
\end{table}

\begin{table}[!htbp]
  \centering
  \caption{\label{Tab4_SM} Same as Table~\ref{Tab1_SM} but for $^{20}$O.}
\begin{ruledtabular}
\begin{tabular}{ccccccc}
  $\tau_c$~[MeV$^{-1}$]     &0.02          & 0.06        &0.10         &0.14        &0.18        &0.22\\
\hline
  $r_{\rm ch}$ [fm]         &$2.801(16)$   &$2.794(18)$  &$2.795(22)$  &$2.804(26)$ &$2.810(32)$ &$2.819(38)$\\
  $r_{m}$ [fm]              &$2.844(12)$   &$2.845(20)$  &$2.847(23)$  &$2.859(28)$ &$2.863(33)$ &$2.883(38)$\\
\end{tabular}
\end{ruledtabular}
\end{table}

%references for SM

\end{document}